\documentclass[6pt, preprint2]{aastex}

\begin{document}

\title{The Peculiar Shape of the $\beta_{app}-z$ Distribution Seen in Radio Loud AGN Jets Is Explained Simply and Naturally In the Local Quasar Model}

\author{M.B. Bell\altaffilmark{1}}

\altaffiltext{1}{Herzberg Institute of Astrophysics,
National Research Council of Canada, 100 Sussex Drive, Ottawa,
ON, Canada K1A 0R6;
morley.bell@nrc.gc.ca}

\begin{abstract}

Recently, it was argued that the log($z$)-m$_{v}$ plot of 106,000 AGN galaxies could be interpreted as an evolutionary path followed by local AGN galaxies as they age. It was suggested that these objects are born as quasars with a high intrinsic redshift component that decreases with time. When the intrinsic component is large it causes them to be pushed above the standard candle line for brightest radio galaxies on a log($z$)-m$_{v}$ plot. In the jets of radio loud AGN galaxies, $\beta_{app}$ is the apparent transverse velocity of the ejected material relative to the speed of light. In the cosmological redshift (CR) model the $\beta_{app}$ vs $z$ distribution has a peculiar shape, and there have been several attempts to explain it. In agreement with the model proposed to explain the log($z$)-m$_{v}$ plot, it is shown here that the peculiar shape of the $\beta_{app}-z$ distribution in the CR model is exactly what is expected \em if the sources are local but their large intrinsic redshifts are assumed to be cosmological in the calculation of $\beta_{app}$. \em This result not only supports our previous interpretation of the log($z$)-m$_{v}$ plot, it further implies that if a large component of the redshift is intrinsic a similar effect should be visible when other parameters are plotted vs $z$. Examining this it is also found that the results are consistent with the local model.


\end{abstract}

\keywords{galaxies: active - galaxies: distances and redshifts - galaxies: quasars: general}

\section{INTRODUCTION}

If quasar redshifts are cosmological (i.e. due to the expansion of the Universe), and are then a measure of the distance to these sources, the angular motions seen in the jets of radio-loud quasars lead, in many cases, to apparent transverse velocities that exceed the speed of light. When quasars were first discovered \citep{mat63,sch63}, some astronomers suggested that they may not be at the distance indicated by their redshifts (see \citet{bur67} for an extensive discussion and early history). Others argued that if quasars were nearby we should expect to detect some proper motion. Soon after, large proper motions were detected in the material being ejected in jets from the centers of these objects, but even when some apparent linear speeds in excess of 30c were measured most astronomers were still reluctant to accept the possibility that quasars might be closer than their redshifts implied. This was the case even though the highly superluminal sources were only found in the sources with high redshifts, which could be taken as a major clue that the explanation was somehow related to the redshifts. However, elaborate relativistic beaming models were soon devised to explain the observations in the CR model \citep[see for example]{bla79}.

The obvious lack of any high-$\beta_{app}$ sources at low and intermediate redshifts has previously been questioned \citep{urr95}, and several attempts have been made by others to explain the peculiar shape of the $\beta_{app}-z$ plot in the CR model \citep{coh88,ver95,kel04,coh06}. However, the results of these investigations have not been completely convincing, and some explanations have been quite invloved. Our purpose here is simply to see how easily this plot can be explained in the local quasar model.

\section{THE DATA}

In this paper quasars may also be referred to as high redshift AGN galaxies.
The $\beta_{app}-z$ plot is shown in Fig 1, which uses the 114 radio loud AGN galaxies with the most accurate angular motion data studied by \citet{kel04}. The highest value observed for the angular motion in each source was used. Not only are high $\beta_{app}$ values missing at low redshifts, there is also a sharp upper cut-off visible at all redshifts, and an abrupt decrease in this cut-off at high redshifts. Here $\beta_{app}$ = $1.58\times10^{-2}$D$_{A}\mu(1+z)$, where D$_{A}$ is the angular size distance to the source in Mpc, $\mu$ is the observed angular motion in mas yr$^{-1}$, and $z$ is the observed redshift. The $\beta_{app}$ values have been taken from \citet{kel04}.

\begin{figure}
\hspace{0.0cm}
\vspace{0.0cm}
\epsscale{1.0}
\includegraphics[width=8cm]{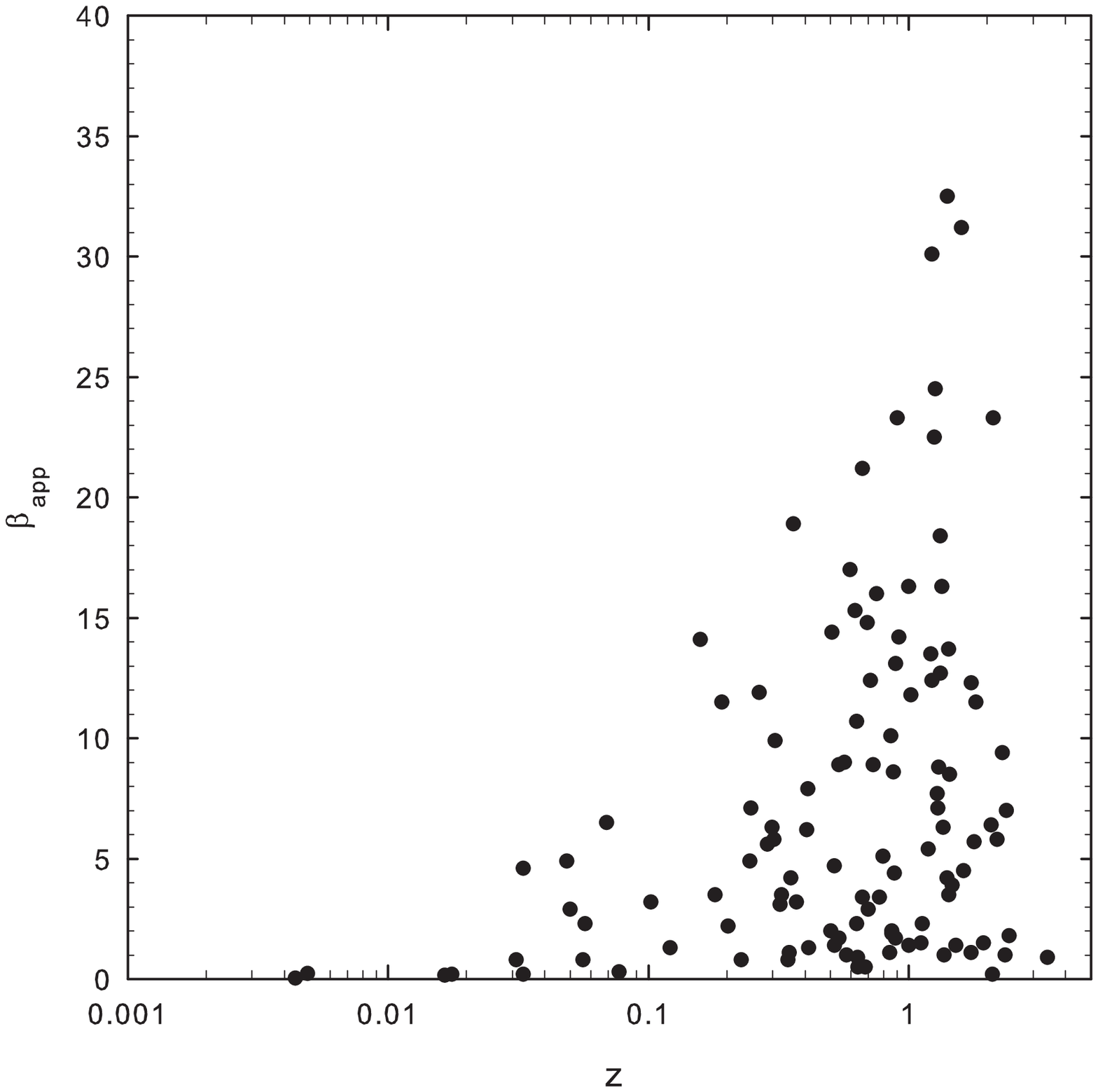}
\caption{{$\beta_{app}$ plotted vs $z$. Data are from \citet{kel04}. \label{fig1}}}
\end{figure}


\begin{figure}
\hspace{-0.8cm}
\vspace{-1.7cm}
\epsscale{1.0}
\includegraphics[width=9cm]{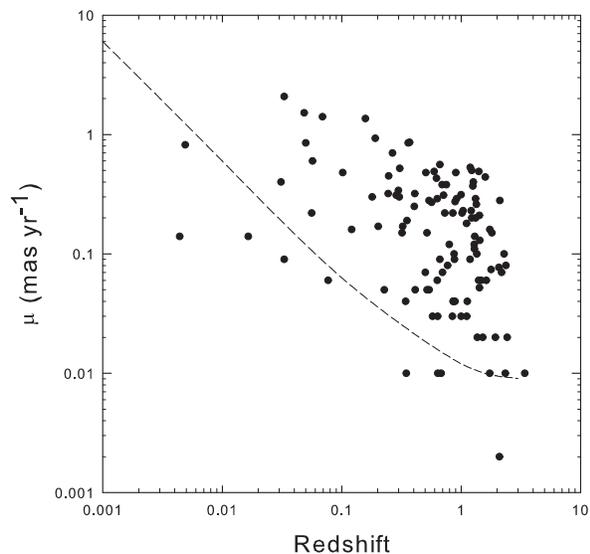}
\caption{{$\mu$ plotted vs $z$. Data are from \citet{kel04}. See text for an explanation of the dashed line. \label{fig2}}}
\end{figure}

One can argue that if the redshifts are a measure of distance the density of sources will decrease to the left in Fig 1 because of the decreasing volume of space. This suggests that if the redshifts are cosmological, few sources will be expected on the left side of this plot. However, this argument does not explain the abrupt cut-off defined by the upper edge of the distribution at all redshifts, its overall shape, or its sharp downturn above $z$ = 1. The reality of the abrupt upper cut-off in this plot is not in question. It was first recognized by \citet{coh88}, and was discussed in more detail by \citet{kel04}, where an attempt was made to explain it as being due to a maximum in the Lorentz factor, $\gamma$, on the $\mu$ vs $z$ plot. \citet{coh06} also argue that it is not due to a selection effect, but to something intrinsic. 

\section{ANALYSIS}

Recently it was argued that the log($z$)-m$_{v}$ plot for 106,000 AGN galaxies \citep{bel07a} could be interpreted as an evolutionary path for AGN galaxies that could be explained if the high redshift AGN galaxies (quasars) were local but were pushed above the standard candle line for brightest radio galaxies by the presence of a large intrinsic component in their redshifts. It has since been pointed out by \citet{arp07} that the log($z$)-m$_{v}$ plot was already exhibiting a similar trend in 1968 when a much smaller sample containing only a few tens of quasars was available. Here we use this local model in an attempt to explain the peculiar shape of the distribution in Fig 1. The model has previously been referred to as the decreasing intrinsic redshift (DIR) model and additional evidence supporting it can be found in \citet{arp07,bel02a,bel02b,bel02c,bel02d,bel04,bel06,bel07b,bcr04,mcd06,mcd07,bur99,chu98,lop06}.

In this type of analysis it is most important to be able to explain the raw data (i.e. the raw observables uncontaminated by modeling). Here parameters such as the angular motion, $\mu$, and the flux density, S, are examples of raw data. Parameters such as $\beta_{app}$, the Lorentz factor, $\gamma$, and the radio luminosity, L, are model dependent, and have meaning only in the model that defines the source distance. Before examining the peculiar shape of the $\beta_{app}$ distribution we first look more closely at which raw observables might influence its shape.

The distribution of $\mu$-values with redshift for these sources is shown in Fig 2 for the fastest and most accurate values obtained at 15 GHz using the VLBA \citep{kel04}. It, like Fig 1, also has a clear upper cut-off. Similarly, when $\mu$ is plotted vs the flux density of the central engine, S$_{15}$, on logarithmic scales in Fig 3, an abrupt upper cut-off is seen. It has a slope of 0.5, and has been discussed previously \citep{mcd07}. This slope is produced naturally by ejections in the local model when ejection is in the plane of the sky, provided the ejection velocity has a maximum value. It has been found that an upper cut-off with a slope of 0.5 is obtained regardless of whether it is $\mu$, $\mu(1+z)$, or $\mu$(1+z)$^{2}$ that is plotted versus S$_{15}$. That this is true indicates that \em the same slope of 0.5 must be present at all redshifts. \em Since the $\mu$ vs S plot, and the $\mu$ vs $z$ plot in Fig 2, both use only the directly observable parameters $\mu$, S and $z$, these plots must be explainable in any acceptable interpretation of $z$. The 19 sources plotted as open circles in Fig 3 have been chosen here to define those sources that fall along the upper cut-off.

It was demonstrated previously that because of the slope of 0.5 on the upper cut-off of the $\mu$ vs S$_{15}$ plot, the upper cut-off on the $\mu$ vs $z$ plot will not be sharply defined \citep{mcd07}. This is true regardless of the cause of the upper slope, or the model assumed. It was also shown that the resulting smearing effect that this slope causes on the $\mu$ vs $z$ plot can be removed in both the DIR and CR models by normalizing all $\mu$-values to a common S$_{15}$ value \citep{mcd07}. When this is done the result is shown in Fig 4 of this paper, and in Fig 3 of \citet{mcd07}. The upper cut-off is more sharply defined in both of these plots, but the improvement is most obvious on the linear plot in Fig 3 of \citet{mcd07}. A smooth decrease is also seen at high redshifts in the log-log plot in Fig 4. It is suggested here that the smoothness of the decrease at redshifts above $z$ = 1 may indicate that it is real, and not just due to some high-$z$ detection cut-off.

The 19 sources plotted as open circles that lie along the upper cut-off in Fig 3 are plotted again as open circles in Fig 4. Clearly the abrupt upper cut-off in Fig 4 is produced by the same sources that produce the upper cut-off in the $\mu$ vs S plot. This means that in the local model the abrupt upper cut-off in the $\mu$ versus $z$ plot can also be explained naturally by the maximum angular motion seen in a simple ejection model \em when ejection is in the plane of the sky. \em

\begin{figure}
\hspace{-1.0cm}
\vspace{-1.6cm}
\epsscale{0.9}
\includegraphics[width=8cm]{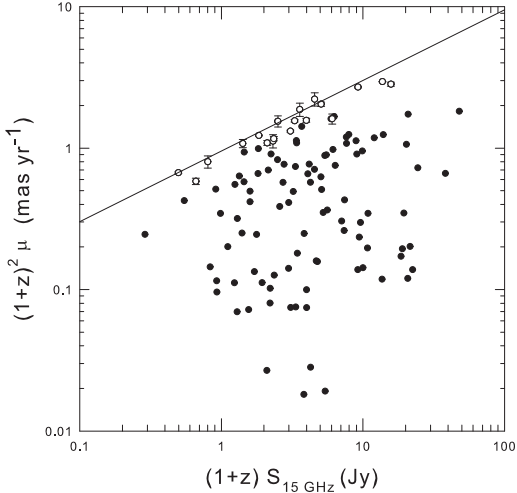}
\caption{{Angular motion vs k-corrected flux density for superluminal sources from \citet{kel04}. See text for an explanation of sources plotted as open circles. \label{fig3}}}
\end{figure}

\begin{figure}
\hspace{-0.8cm} 
\vspace{-2.3cm}
\epsscale{0.9}
\includegraphics[width=8cm]{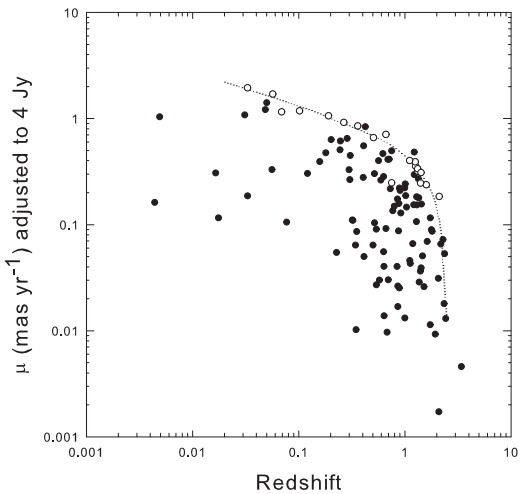}
\caption{{Same as Fig 2 after normalizing $\mu$ values to S = 4 Jy. The 19 sources plotted as open circles are the same as those plotted in Fig 3.  \label{fig4}}}
\end{figure}

\section{THE PECULIAR SHAPE OF THE $\beta_{app}-z$ PLOT}

We now need to determine what produces the peculiar shape of the $\beta_{app}$ distribution in Fig 1 and its abrupt upper cut-off. In the local model, where the central engine has been found to be a good radio standard candle, the radio-loud sources studied here will represent the closest. Those found in optical surveys (e.g. the SDSS) can be located considerably further away. In any event, the value assumed for the distance component of the redshift, $z_{c}$, is not critical so long as it is relatively small. The remainder of the observed redshift, $z$, is then due to an intrinsic component, $z_{i}$, where $(1+z) = (1+z_{c})(1+z_{i})$. For small $z_{c}$ values the intrinsic redshift component is essentially equal to the observed redshift. Using the relation $\beta_{app}$ = 6.6x10$\mu$D$_{z}$(1+$z_{c}$), we now recalculate $\beta_{app}$. Here $\mu$ is in mas yr$^{-1}$, D$_{z}$ is the redshift distance in units of redshift calculated using the appropriate cosmology calculator from the NASA/IPAC Extragalactic Database website with H$_{o}$ = 70, $\Omega_{M}$ = 0.3 and $\Omega_{L}$ = 0.7. These constants have been used to keep the calculations as consistent as possible with those of \citet{kel04}.

\begin{figure}
\hspace{-0.5cm}
\vspace{-1.9cm}
\epsscale{0.9}
\includegraphics[width=8cm]{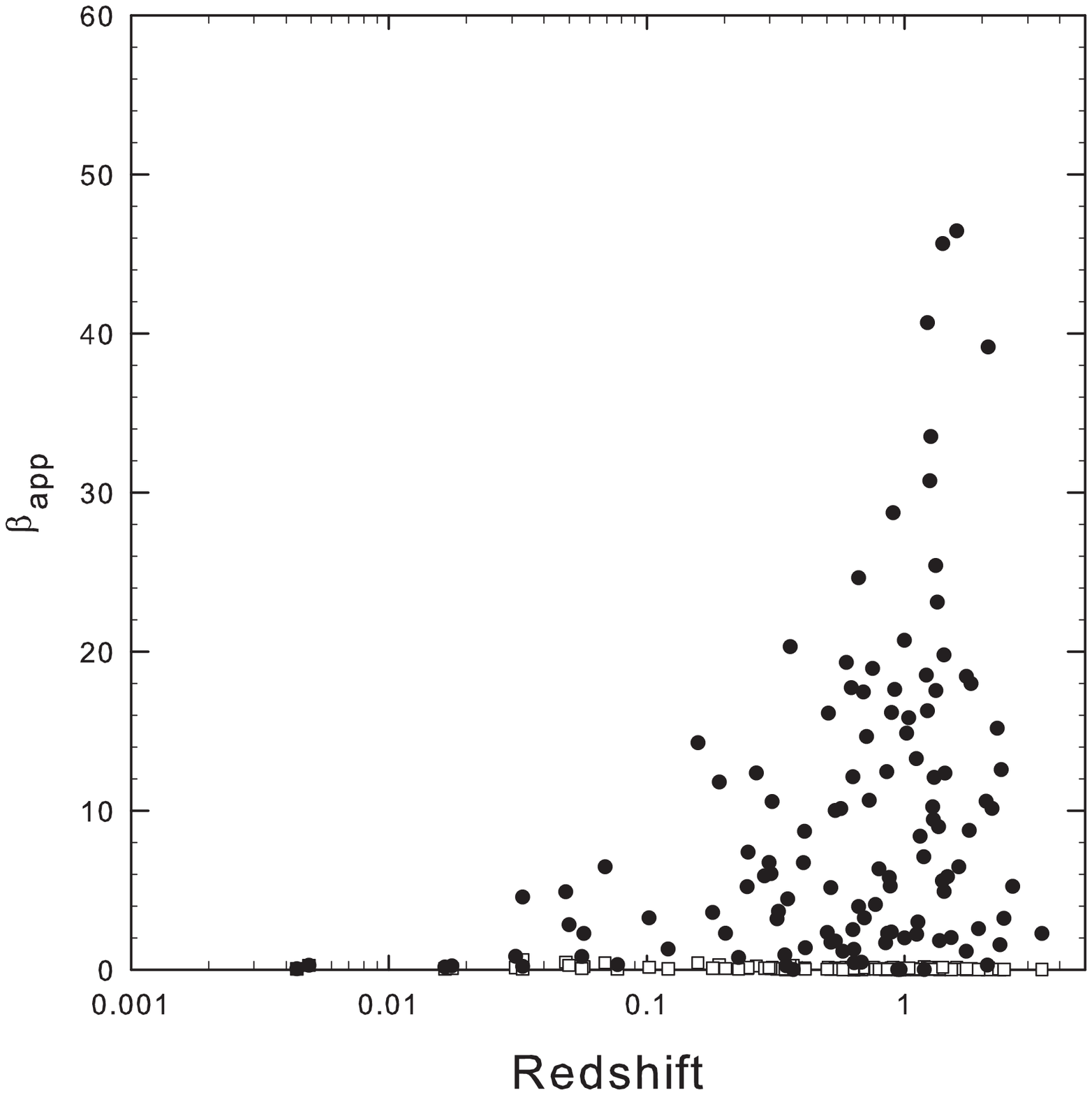}
\caption{{(open squares) $\beta_{app}$ calculated as described in the text using raw $\mu$ values and D$_{z}$ = $z_{c}$ = 0.005. (filled circles) $\beta_{app}$ calculated using raw $\mu$ values and D$_{z}$ = $z$.\label{fig5}}} 
\end{figure}

\begin{figure}
\hspace{-0.5cm}
\vspace{-2.0cm}
\epsscale{0.9}
\includegraphics[width=8cm]{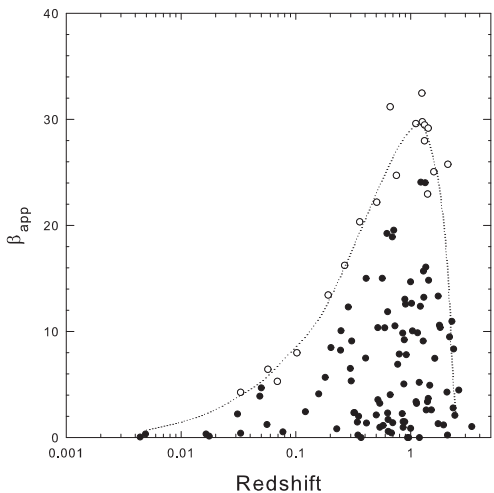}
\caption{{Same as Fig 4 except here $\mu$ values have been normalized to S$_{15}$ = 4 Jy. See text for a description of how the dotted curve was generated. \label{fig6}}} 
\end{figure}

When $z_{c} = 0.005$ is used in the calculation of D$_{z}$, the result is plotted versus $z$ as open squares in Fig 5. When $z$ is used in the calculation of D$_{z}$ the resulting $\beta_{app}$ values are plotted as filled circles in Fig 5. It is immediately obvious that when the intrinsic component is included in the D$_{z}$ calculation, the shape of the distribution in Fig 5 becomes the same as that seen in Fig 1. \em This indicates that a $\beta_{app}$ distribution with the same shape as the distribution in Fig 1 will result if the sources are all relatively nearby but there is a large intrinsic component in the redshift which is included, incorrectly, in the calculation of D$_{z}$ when determining $\beta_{app}$. \em

It is easily shown that the peculiar shape of the $\beta_{app}-z$ distribution is not significantly affected by either the choice of cosmology or by the method used to calculate the distance (i.e. redshift distance, luminosity distance, or angular size distance). For the observed $\mu$-values, the shape of the $\beta_{app}-z$ distribution is almost entirely based simply on whether the redshift used in calculating D$_{z}$ is $z_{c}$ or $z$, where $z$ contains both $z_{c}$ and $z_{i}$.

In Fig 6 the $\beta_{app}$ values in Fig 5 have been recalculated after normalizing $\mu$ to S = 4Jy as was done in Fig 4. Using the $\mu$-values defined by the dotted line in Fig 4 we also calculated the maximum $\beta_{app}$ values, as a function of $z$, that correspond to these maximum $\mu$ values. These are shown by the dotted line in Fig 6. It is thus clear that the upper cut-off of all plots is defined by the maximum observed $\mu$ value at each redshift or S value. In Fig 6, the sources that define the upper limit in Fig 3 are plotted as open circles. It is now also clear that in each plot where an upper cut-off is seen, it can be explained naturally in the local model by the maximum angular motions seen when the direction of ejection is in the plane of the sky.

In the local model the decrease in $\mu$ with increasing $z$, shown by the dotted line in Fig 4, is attributed to a decrease in the ejection speed with increasing intrinsic redshift (towards younger, less luminous sources). Because the upper limit is produced by the maximum $\mu$ at each redshift, this confirms that the upper cut-off in the $\beta_{app}$ distribution in Fig 1 is produced by something intrinsic to the sources (in this case the actual ejection speed, which is proportional to $\mu$). This is in agreement with what was claimed by \citet{coh06}, even though nothing in their model is directly relevant to the DIR model. Unlike in the local model, where the change in the maximum $\mu$ with $z$ is due to a change in ejection speed, in the relativistic beaming model the change in maximum $\mu$ is related to a change in the viewing angle, with the angular motion $\mu$ increasing with $z$ as the viewing angle approaches zero (gets nearer to the line-of-sight). This is a significant difference between these two models, and one which might eventually help in determining which is correct.

 Although \citet{coh06} plotted $\beta_{app}$ versus radio luminosity (L), since L is proportional to $z$ in the CR model, the shape of the upper $\beta_{app}$ cut-off will be the same as in Fig 1. This can be seen in Fig 7 where $\beta_{app}$ is plotted vs L. Here L has been calculated using the peak flux density values listed in \citet{kel04} and therefore may differ slightly from Fig 6 of \citet{coh06}, where the mean flux density was used. Also, here the luminosity distance has been used in the calculation of $\beta_{app}$. The open circles are again the sources defined by open circles in Fig 3.


\section{DISCUSSION}

It has been demonstrated that the abrupt upper cut-off visible in the above plots is produced by the same sources in all cases, and is due to a maximum in the angular motion $\mu$. In the local model these maximum $\mu$ values are produced geometrically when ejection is in the plane of the sky. It has also been shown that the peculiar shape obtained for the $\beta_{app}$ vs $z$ and $\beta_{app}$ vs L distributions in the CR model can be explained simply and naturally in the local model if the large intrinsic component in their redshifts is assumed to be distance related in calculating D$_{z}$.

How easily is the shape of the $\beta_{app}$ distribution explained if the redshifts really are entirely cosmological? It is not claimed here that the explanation for the shape of the distribution put forth by \citet{coh06} cannot be correct. Although, with the number of parameters available in the relativistic beaming model there should be little surprise that it might also be possible to explain the observations in that model. On the $\beta_{app}$ vs L plot it is reasonable to suggest that the high $\beta_{app}$ values will be found in the sources with highest luminosity if the high luminosity is due to Doppler boosting. The increase in  $\beta_{app}$ with L along the upper cut-off in Fig 6 can then be explained if the viewing angle becomes much smaller, approaching the line-of-sight, at high redshifts. Although it may seem reasonable that $\beta_{app}$ could be tied to L in models where Doppler boosting plays a role, it is less obvious why it would be so tightly correlated with redshift. It would be tempting to argue that high $\beta_{app}$ sources are only seen at high redshifts because only highly Doppler boosted sources can be seen at such large distances. However, many low-$\beta_{app}$ source are also detected at high redshifts and these are unlikely to be strongly boosted. We might also wonder, with so many jets pointed towards us at high redshift, why there is not even one at low redshift. 

A weak part of the arguments presented by \citet{coh06} would also seem to be that they draw on the many available parameters in various ways, as needed to explain the observations. For example, to explain the low-speed quasar jets the \em pattern speeds \em must be substantially slower than the \em beam speeds. \em However, for the jets with the fastest speeds, \em pattern \em and \em beam \em speeds are required to be approximately equal. This appears somewhat arbitrary and contrived. These investigators also have problems explaining other observations. For example, because the lobes of Cygnus A are so powerful the jets are expected to be highly relativistic. They are not, and this requires the introduction of the two-component beam model (fast spine and slow sheath). Similarly, the asymmetric beams in M87 are unexpectedly found in $\lambda$-2cm VLBA monitoring to be non-relativistic \citep{kov07,bel07b}, contrary to the predictions of the relativistic beaming model. M87 is one of the few sources whose distance is accurately known and where the relativistic beaming model can then be tested. It appears to have failed the test. But then this comes back to the basic question being asked here concerning Fig 1 of why the low redshift sources (the ones whose distances are most accurately known) do not exhibit highly relativistic motions at $\lambda$-2cm. This may suggest that there are no highly relativistic motions at high redshifts either, and that they have just been created by assuming that a large intrinsic component in the high redshift sources is distance related, as concluded above.

Most will also agree that the discussion of the so-called \em inversion problem \em in sections 2 to 4 of \citet{coh06} is considerably more complex than anything required in the local model to explain the above plots.
In fact, explaining the $\beta_{app}$ distribution curve in the CR model is, in almost every case, much more complicated than the simple explanation we have presented here to explain it in the local model.

\begin{figure}
\hspace{-0.5cm}
\vspace{-2.0cm}
\epsscale{0.9}
\includegraphics[width=8.5cm]{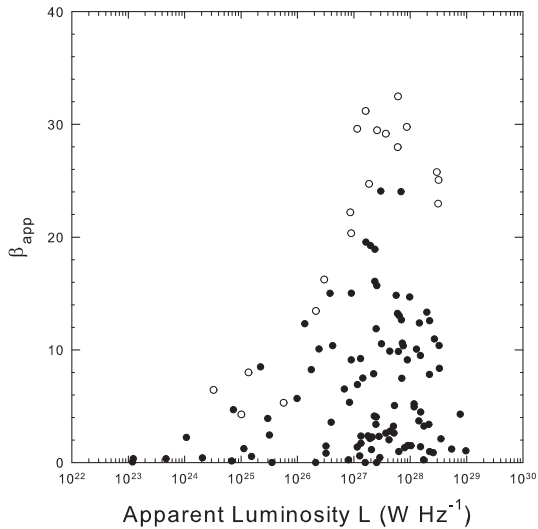}
\caption{{$\beta_{app}$ after $\mu$ normalized to 4 Jy plotted vs Luminosity L. Sources plotted as open circles are the same as in Fig 6. \label{fig7}}}
\end{figure} 
   
\begin{figure}
\hspace{-0.5cm}
\vspace{-2.8cm}
\epsscale{0.9}
\includegraphics[width=8.5cm]{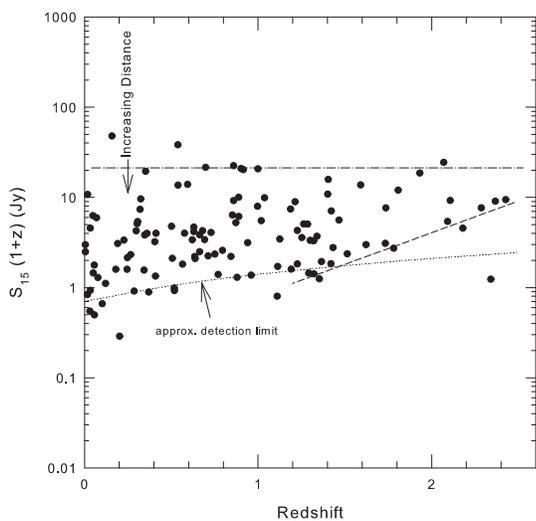}
\caption{{K-corrected S$_{15}$ plotted vs redshift. See text for an explanation of the dashed and dotted lines. \label{fig8}}}
\end{figure}

\subsection{Is there Other Evidence in the Raw Data that Supports the Local Model?}

If there is a large intrinsic component in the redshifts that is always assumed in the CR model to be distance related it should produce a similar effect when other parameters are plotted vs redshift. In Fig 2, where $\mu$ is plotted vs redshift, the dashed line shows how the maximum angular motion would fall off in a de Sitter cosmology if the ejection speed were constant. Although most sources are located at much larger redshifts than would be expected for a constant $\mu$ in the CR model, this is again easily explained if their redshifts contain a large intrinsic component that pushes them to higher redshifts, in the same way the sources get pushed above the standard candle slope on the log($z$)-m$_{v}$ plot \citep{bel07a}, and $\beta_{app}$ gets pushed up at high $z$ in Figs 1 and 5.

In Fig 8, the k-corrected flux density for the radio loud AGN galaxies from \citet{kel04} is plotted versus $z$. In this plot the dotted line represents the approximate detection limit, and the dashed line at high redshifts is explained in the DIR model by the plate cut-off \citep{bel06}. It has not been explained in the CR model, but shows up in all early radio source surveys, except when a special effort was made to extend the plate cut-off to more sensitive levels \citep[see Figs 4, 6, 16, 17, 18, and 19 of that paper]{bel06}. The upper cut-off in the source distribution (upper dashed line) is expected to be quite abrupt in the local model where it is interpreted as being due to a rapid reduction in source numbers as the volume of space sampled decreases. As can be seen, instead of falling off steeply with redshift, as would be expected if the sources were standard candles in the CR model, the upper cut-off is quite flat. This zero slope is required if the sources are good standard candles in the DIR model as is suggested by the slope of 0.5 found previously for the upper cut-off \citep{mcd07}, and here in Fig 3. The location of the high redshift sources in Fig 8 can then easily be explained when the redshifts contain a large intrinsic component. However, if the redshifts are entirely cosmological, a zero slope would have to have happened purely by chance.

\subsection{Some Difficulties with the Local Model}

Although many raw data relations have been shown to support the local model, there are also some things that are less easily explained in the local model than in the CR model. It is true that the intrinsic redshifts themselves have not been explained, but we need to remember that Dark Matter and Dark Energy also have not been explained. The difference here would seem to be that we can accept these without explanation because they are needed to explain the observations in the currently accepted model, whereas intrinsic redshifts are not. However, the lensing of some quasars by intervening galaxies may also prove difficult to explain unless these quasars also have a significant distance component and have been detected simply because their radiation has been intensified by the lensing.

Perhaps even more difficult to explain in the local model is the evidence for absorption at a high redshift in the spectrum of a quasar that, in the local model, must be relative nearby. It would seem that the only way to explain this would be if the intervening absorbing gas itself has a large intrinsic component. In other words the absorber itself must be a quasar and both could be relatively local. However, the chance of this happening may not be that small if many objects are ejected, at different times, from a parent active galaxy. Since the intrinsic component decreases with time in the DIR model, to be located between us and another quasar of higher intrinsic redshift, the intervening material would have to be further from the parent. It would then naturally be older and, in the local model, would then have a smaller intrinsic component. In fact, \citet{bow06} have already reported detecting QSO absorption lines from other intervening QSOs. Also, until the mechanism for producing an intrinsic component is known, defining a scenario to explain the observations cannot always be expected to be straight-forward. It would seem to be most important then that all the evidence be taken into account before any conclusions about intrinsic redshifts are drawn.

It is also worth noting that a lot of work has been done on damped Ly$\alpha$ absorption lines, seen towards QSOs, whose redshifts cover a large range. However, as pointed out by \citet{yor07}, the metallicity of the absorbers shows very little evolution between $z \sim$ 3 and $z \sim$ 0. This runs contrary to the expectation that the metallicity should rise towards lower redshifts if the DLAs trace the bulk of gas in galaxies. On the other hand, there is no problem explaining this if the sources are all nearby. Unfortunately, our understanding of the basic properties of the absorbing material, such as its mass, size, star formation rates and luminosity, still remains very limited \citep{wol05,yor07}. 

It is important to remember that there is much evidence that is more easily explained in the local model than in the cosmological one, such as that discussed here.  Moreover, since in the DIR model the proposed intrinsic component is only present for the first $\sim1\%$ of the life of a galaxy, intrinsic redshifts in quasars no longer need to threaten generally accepted models like the Big Bang, inflation, or the cosmological redshift model for normal galaxies.

\section{CONCLUSIONS}

 In the local model most high redshift quasars detected to date are assumed to have a large intrinsic redshift component superimposed on a small cosmological one. It has been shown here that the peculiar shape of the observed $\beta_{app}$ distribution is obtained naturally in this model when the total redshift is used, incorrectly, in the calculation of $\beta_{app}$. This, in combination with the measured $\mu$-values, produces a peak in the $\beta_{app}$ distribution at high redshifts the same way AGN galaxies with high redshift get pushed above the bright edge of the radio galaxy standard candle line on a log($z$)-m$_{v}$ plot. In this simple model the abrupt upper cut-off in $\beta_{app}$ is also explained naturally by a geometrically produced maximum in the angular motions, $\mu$, which results when ejections are in the plane of the sky. The order-of-magnitude increase in the maximum $\mu$ with decreasing intrinsic redshift (towards the more mature radio galaxies) can be explained by an increase in the ejection speeds as the sources become more luminous. It has also been demonstrated that in addition to explaining the $\beta_{app}-z$ and log($z$)-m$_{v}$ plots, the presence of an intrinsic redshift component also nicely explains the shape of the $\mu-z$ and S-$z$ plots. In most respects this model is much simpler than what is required to explain the observations in the CR model. While there is no difficulty explaining why there are no high $\beta_{app}$ sources at low redshifts in the local model, \em this still remains to be explained in the cosmological redshift model. \em

\section{ACKNOWLEDGEMENTS}

This research has made use of the NASA/IPAC Extragalactic Database (NED)
which is operated by the Jet Propulsion Laboratory, California Institute of Technology, under contract with the National Aeronautics and Space Administration.


\end{document}